\documentclass{article}
\pdfoutput=1

\usepackage{arxiv}

\usepackage{graphicx}
\usepackage{lineno}
\usepackage[utf8]{inputenc} 
\usepackage[T1]{fontenc}    
\usepackage{hyperref}       
\usepackage{url}            
\usepackage{booktabs}       
\usepackage{amsfonts}       
\usepackage{nicefrac}       
\usepackage{microtype}      
\usepackage{lipsum}

\title{Autonomous Industrial Management via Reinforcement Learning: Self-Learning Agents for Decision-Making -- A Review}

\author{Leonardo A. Espinosa Leal\thanks{Corresponding author: leonardo.espinosaleal@arcada.fi}\\
Department of Business Management and Analytics \\ Arcada University of Applied Sciences\\ Helsinki, Finland
\And
Magnus Westerlund \\
Department of Business Management and Analytics \\ Arcada University of Applied Sciences\\ Helsinki, Finland
\And
Anthony Chapman \\
Computing Department\\ University of Aberdeen\\
Aberdeen, UK.}

\begin{document}
\maketitle

\begin{abstract}
Industry has always been in the pursuit of becoming more economically efficient and the current focus has been to reduce human labour using modern technologies. Even with cutting edge technologies, which range from packaging robots to AI for fault detection, there is still some ambiguity on the aims of some new systems, namely, whether they are automated or autonomous. In this paper we indicate the distinctions between automated and autonomous system as well as review the current literature and identify the core challenges for creating learning mechanisms of autonomous agents. We discuss using different types of extended realities, such as digital twins, to train reinforcement learning agents to learn specific tasks through generalization. Once generalization is achieved, we discuss how these can be used to develop self-learning agents. We then introduce self-play scenarios and how they can be used to teach self-learning agents through a supportive environment which focuses on how the agents can adapt to different real-world environments. 
\end{abstract}


\section{Introduction}

Recent developments in industry, such as Industry4.0, have emphasized the need for a comprehensive automation of operational processes~\cite{schwab2017fourth}. Similar developments in both the automobile~\cite{Tesla} and maritime industry~\cite{Rolls-Royce} has highlighted the need for vehicle/vessel driving automation. In recent media, we often see the term autonomous used as a synonym for automated. Hence, automated driving often become autonomous driving, without a deeper reflection on what an autonomous system would entail. Within information systems research we also have a long history of researching autonomous entities, sometimes as an abstract futuristic entity that is often ill-defined in terms of true autonomy, although some efforts have been made \cite{franklin1996agent}. We consider an autonomous entity as an evolutionary leap compared to an automated entity. Meanwhile, the automation component is far from solved for many physical environments. On the other hand, for digital environments such as stock exchanges and advertisement auctions, the process has, to a large extent, already been automated, but autonomous behaviour is rarely needed or desired in such environments. To address this gap we delimit our focus in this paper, to consider one part of an autonomous system, the learning mechanisms of an autonomous entity. To achieve this, we first expand on the subject of what an autonomous system entails and relevant areas that may assist in achieving autonomy.

The need for automation has led to a realization that we also need to digitize our industrial environments and processes. These processes include manufacturing, warehousing, building management, or hospital management. Digitization may happen by introducing IoT sensors, communication networks, and cameras into an environment and then digitally capture the physical reality. Advanced types of automation will often form a model using past events and test the model using more recent events in a controlled environment before testing the model in real scenarios. A common approach has been to use fuzzy logic to describe expert behaviour for an entity or an agent that operates in or oversees the environment \cite{carlsson2012fuzzy}. Prediction is sometimes required in automation and neural networks are a popular way to compare on the outcomes of the predictions \cite{wang2016combined}. 

In order to verify a number of algoritmically derived scenarios in the physical factory may often be expensive, dangerous, and disruptive. Instead a digital representation of the process/environment, a so called extended reality, can be used for verification and optimization, before ideas/changes are moved into the physical realm. An example of an extended reality is the digital twin technology \cite{boschert2018next}, which can be described in many formats, e.g. process graphs, a 2D flat model, a 3D space, or a 4D space-time environment (can be seen as stacked 3D models with temporal dependencies). Although, the more realistic the environment and simulation is, the better the final result can be expected. Complexity tend to increase for each added dimension and different strategies must be devised to conquer the different environments. 

A state-of-the-art form of digital twin technology is physics-based (follows physical laws) and allows the agent to interact with the environment, i.e. bi-directional data-exchange \cite{kritzinger2018digital}. Furthermore, the environment may be also be modified within certain parameters to allow for an improved process. Hence, both the agent and the extended reality environment are part of the modelling process and may be enhanced in order to improve an outcome. In this article, we focus on a method and experimental study for self-playing agents which can use deep reinforcement learning, in what can best be generalized as an extended reality environment \cite{espinosa2019reinforcement}. Our aim is to address the topic of designing self-play scenarios for self-learning agents that combine various sources for discovering and generating training data. By reviewing the literature we noticed a gap which could improve existing architectures used for creating self-learning agents that can be used for enabling autonomous industrial management. \cite{kritzinger2018digital} also highlight the need for more concrete digital twin case-studies. Combining AI research and digital twin technology offer an interesting avenue for further research.

The structure of the paper is the following. Section 2 reflect on three core terms used throughout the paper. The first subsection provides a discussion on autonomous vs. automated systems, and notes that although media and information systems literature provide a considerable history of autonomous systems, real-world examples of such systems do not yet exist in a strict language or legal sense. Following this, eXtended Reality (XR) is introduced as a self-play environment for teaching models through interactivity. The last sub-section discusses the need for data augmentation to bridge the gap between training in extended reality and inference in the physical reality. The third section introduces reinforcement learning and recent advancements. The fourth section presents how algorithms can learn by self-play, i.e. learning from the interaction in XR, and a conceptual architecture. The fifth section contributes a design and proposes applications in some relevant areas. The final section concludes our findings and presents future directions for research.
\section{Research methods and concepts}\label{research_methods}
In the first part of this paper, a literature review of the relevant concepts is presented. Through the literature review we provide a foundational understanding of root definitions for the second part, where we provide an analysis of our findings. In this latter part, we follow a soft systems methodology (SSM) by framing a problem formulation (model learning) and an action plan (conceptual model for learning) aimed at future research \cite{checkland2013soft}. \cite{sorensen2010conceptual} state that SSM consist of an analysis of the current status of the system. This includes inherent problems and activities, and a definition of the system that derives the actual goal of the targeted system (“root definition”) in order to propose a conceptual system model.

\subsection{Requirements for achieving autonomous systems}\label{section21}
Autonomous systems have long been a promise of both academia and industry alike. One of the early academic papers discussing autonomous systems was presented in the 1950s, where the author discusses an autonomous nonlinear oscillation system \cite{rauch1950ii}. Much later, in the 1990's with the introduction of the personal computer and precursors to the Internet, autonomous agents became a topic of interest. The agent implementations tended to cover relatively narrow and software localized areas. \cite{smith1994kidsim} defines the agent as a \emph{persistent software entity dedicated to a specific purpose}. \cite{franklin1996agent} reviews the early definitions of autonomous agents, and then summarizes that the autonomous property can be defined as \emph{something which exercises control over its own actions}. Given the progress since, we would argue these definitions do not qualify an entity for being autonomous, but mere automated.

The definition of the word autonomous, in a language sense, can be describe as \emph{an autonomous entity is independent and has the freedom to govern itself} (adapted from the Cambridge Online Dictionary). Using a strict interpretation, it suggests that truly autonomous systems have yet to come into existence, and that traditional methods have not been able to deliver beyond the automation component. Automation can here be defined as \emph{an entity made to operate by the use of machines or computers in order to reduce the work done by humans} (adapted from the Cambridge Online Dictionary). This latter definition of automated systems corresponds better to the original definition of autonomous agents, as presented above. An autonomous system should have the freedom to govern itself, and governance implies some form of understanding of real-world repercussions of ones actions, for example the limits the legal framework provide us with. Additionally, an autonomous entity should show independence, suggests that a mere exploitation of the historical world is not sufficient, but rather that the entity must learn to explore unknown worlds through self-play. Hence, an entity that employs an intelligence derived from a rule-set or a neural network trained on historical data, will likely fail the test of true autonomy.

A similar standpoint has been taken by organizations regulating and standardizing the automotive industry \cite{kyriakidis2015public}. The Society of Automotive Engineers (SAE) through the On-Road Automated Vehicle Standards Committee is  responsible for defining levels of driving automation for cars and provides a classification of levels from 0-5. In the SAE classification, the highest level, "Level 5 (steering wheel optional)" is defined as a fully automated sysem, and not as an autonomous system \cite{sae2014taxonomy}. 

\begin{figure}
\begin{center}
\includegraphics[width=0.3\paperwidth]{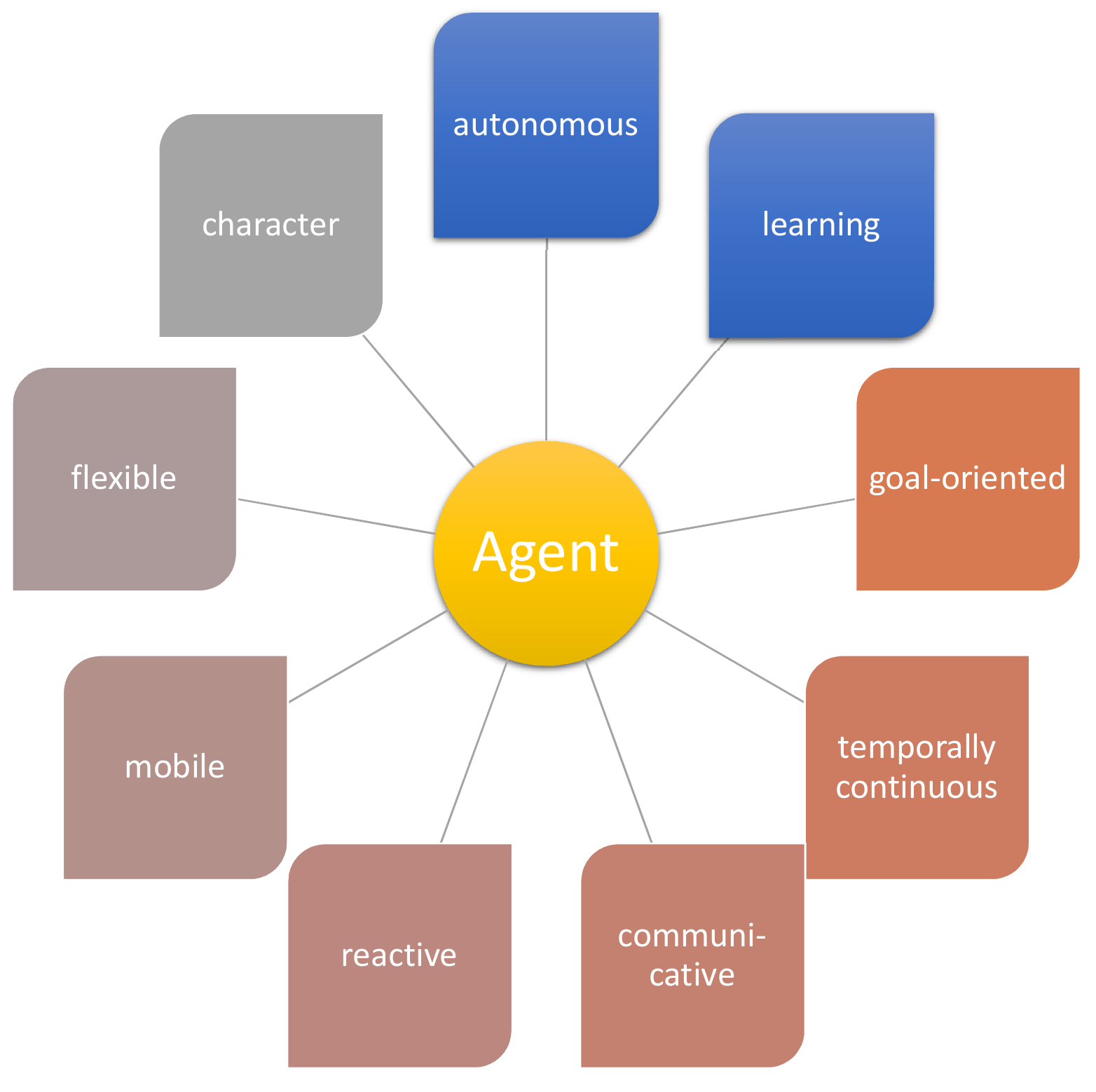}
\caption{Agent properties adapted from \cite{franklin1996agent}.}
\label{fig0}    
\end{center}
\end{figure}

There have been several methods proposed for creating agents, e.g. fuzzy logic or neural networks. In \cite{carlsson2012fuzzy} the authors use the concept of intelligent agents for describing an important property of the entity. Intelligence or rather reasoning, allows an agent to react to different defined circumstances or even unforeseen circumstances. Some early considerations realized the challenge of defining a globally validated view of a contained entity, with so called artificial intelligence, and instead highlighted that at the essence of an intelligent entity is an ability to perform subjective probability analysis over both logical and natural (physical) conditions, \cite{anscombe1963definition} and \cite{carlsson2012fuzzy} approach this through an expert system governed by fuzzy logic and highlight certain abilities an agent must achieve to be considered intelligent. They highlight certain abilities an agent must achieve to be considered intelligent. For example, an agent must be able to plan a set of actions and needs to adapt its plan to changing conditions. \cite{franklin1996agent} provide a classification of properties an agent may have (see Fig.~\ref{fig0}).

On a more philosophical note, for an autonomous entity to achieve independence and freedom to govern itself, one can argue only an AI with consciousness can achieve such a feat. However, \cite{bach2019consciousness} highlights that consciousness, although not properly understood, may indeed be a representation rather than an agency. Hence, the consciousness created in our minds are a result of the brain attempting to make sense of sensory inputs, and consciousness is but an illusion, a model of the contents of our attention. Therefore, the state of dreaming and the state of awakeness is a change in sensory details and not a termination/commencement of conscious experience. Assuming that the cortical conductor theory interpretation is correct  \cite{bach2018cortical}, we can reduce an autonomous system into four main concepts that feeds an attention layer, learning, reasoning, control, and selection. In this paper we delimit our focus to the learning mechanism that occur from an interaction and describe some industrial environments were learning can occur. Our position is that without an interactive learning mechanism and an ability to self-improve, such that reinforcement learning provide us with, the system cannot be considered autonomous. Still, as discussed above there are other properties that must also be fulfilled for a system to be considered truly autonomous. 

\subsection{Extended reality}
The term eXtended Reality (XR) has become known as a common term for fields were digitally enhanced environments and human-interaction are combined for various purposes. This includes Virtual Reality (VR), Mixed Reality (MR), and Augmented Reality (AR). Compared to, for example, autonomous driving, these environments offer AI researchers an important and relatively low-cost setting for implementing human-interacting algorithms, such algorithms may contain reasoning that can be considered ``true'' AI. Perhaps most importantly, these digital environments allow us to study an agent’s decision making, and thereby, offering a feed-back loop between the agent-environment-user.

Still, the need for training new abilities often require that agents are presented with big data. The traditional approach was often to gather the data from users, e.g. playing a game, and then train on this data. Today, the deep networks' need for massive data and relatively complex training scenarios for reinforcement learning presents researchers with a problem that is often better solved by augmenting additional data for training networks, than using real data. Data augmentation methods depend on the problem at hand, an agent may for example have to learn how to deal with object recognition, spatial actions to take in relation to detected objects, or temporal differences in scenarios, to name a few. The following sub-section reviews some of the relevant literature regarding the performance of data augmentation for training agents.

\subsection{Data augmentation}
Data augmentation has gained prominence as a studied method for extending available datasets \cite{krizhevsky2012imagenet}. The ability to train deep networks often depends on the availability of big data. The fields of both image recognition and voice recognition has been strongly influenced by deep learning methods~\cite{lecun2015deep,Leal2018Web}, and this has motivated the emergence of data augmentation. For RL many of the same concepts can be utilized, but there is also a need for methods that work particularly in the temporal dimension.

Traditional, na\"ive approaches tend to manipulate the investigated environment or dataset in various ways. For visual tasks, these have included scaling of objects, translating i.e. moving objects spatially to various positions, rotation of objects at various angles, flipping objects as to remove bias from any direction, adding noise, changing lightning conditions, and transforming perspective of a known object by changing the angle of view \cite{krizhevsky2012imagenet, pai2017augmentation}.

For audio tasks, data augmentation often includes deformations into a temporal dimension. Approaches include time stretching by changing audio speed, pitch shifting by raising or lowering the tone frequency by various degrees, dynamic range compression, and introducing background noise using gaussian or natural noise methods \cite{salamon2017deep}.

The na\"ive data augmentation approaches tend to produce limited alternative data for RL agents to learn from in an extended reality setting. For an RL agent to learn new abilities, data augmentation must support the agent’s scenario learning process. As suggested by \cite{doya2000reinforcement}, we shift from learning to generalize on spatial data to reacting to continuous-time dynamical systems without a-priori discretization of time, state, and action. 

Several approaches exist for the creation of these scenarios. An important method is adversarial learning, as it can produce new and complex augmented datasets by pitting a generative model against an adversary \cite{goodfellow2014generative}. A generative model in combination with XR, can also address the exploration problem, as exploring some states in the physical reality could be very costly and dangerous. This combination also allows the system developer to understand which state spaces in the virtual environment has been visited and trained upon, and the model's ability to generalize in the extended reality environment. These generative techniques for extending the learning environment are further explored in the following section.

\section{Reinforcement learning}
Although XR is slowly receiving more recognition, AI and machine learning's use of XR to enhance learning is lagging behind. XR could help improve an AI's behaviour by providing information from either pure virtual or semi-real environments. Reinforcement learning is one machine learning technique which, when combined with XR, could produce interesting and beneficial results for many  applications, such as driverless cars, autonomous factories, smart cities, gaming and more. 

RL's primary purpose is to calculate the best action an agent should take when an environment is provided. With RL, we could be able to calculate what best action to take by maximizing the cumulative reward from previous actions, thus learning a policy. RL has long roots from areas such as dynamical programming \cite{bellman1957s}, and for a historical review see \cite{kaelbling1996reinforcement}. However, recently it has received a lot of attention for its potential to advance and improve applications in gaming, robotics, natural language processing (including dialogue systems, machine translation, and text generation), computer vision, neural architecture design, business management, finance, healthcare, intelligent transportation systems, and other computer systems \cite{li2017deep}. 

Attention and memory are two parts from RL which, if done impetuously, could negatively affect performance. Attention is the mechanism which focuses on the salient parts. Whereas, memory provides long term data storage, and attention is an approach for memory addressing \cite{li2017deep}. Using XR and self-play, agents may be able to learn desired behaviour before an action an agent makes become crucial to their performance. As an example, autonomous helicopter software could learn fundamental mechanisms for flight using virtual data in simulations in order to achieve high level of attention using the memory required, without the risks posed by real world applications. Once the attention has reached a desired level, it can be applied to real agents in the physical world. 

General value functions can be used to represent knowledge. RL, arguably, mimics knowledge in the sense that it (generally) learns from the results of actions taken. Thus, one may be able to represent knowledge with general value functions using policies, termination functions, reward functions, and terminal reward functions as parameters \cite{sutton2011horde}. Doing so, an agent may be able to predict the values of sensors, and policies to maximize those sensor values, and answer predictive or goal-oriented questions.

Generative Adversarial Networks (GANs) estimate generative models via an adversarial process by training two models simultaneously \cite{goodfellow2014generative}, a generative model G to capture the data distribution, and a discriminative model D to estimate the probability that a sample comes from the training data but not the generative model G. Such an approach could be extended to XR by training a generative model G on virtual / simulated test data and then a discriminative model D to estimate the probability that a sample comes from the real world. This could help tackle some of the issues with RL within virtual environments and extended to the real world. RL and XR could be used before the agent is applied to a real environment, this could save on resources and make autonomous systems a more viable option for general use.  

GANs together with transfer learning could advance self-play using virtual environments for real world agents \cite{pan2010survey}. By combining virtual data generative models and transferring the learning model to a discriminative model, we may be able to accurately express what was learned from the virtual learning environment to the real agent. Again, unforeseen problem will inevitably arise due to the nature of modelling. By using RL both in the virtual learning phase and embedded into the real agent, we may drastically improve a real agent's learning time. 

\cite{vezhnevets2016strategic} proposed a strategic attentive writer (STRAW), a deep recurrent neural network architecture, for learning high-level temporally abstracted macro-actions in an end-to-end manner based on observations from the environment. Macro-actions are sequences of actions commonly occurring. STRAW builds a multi-step action plan, updated periodically based on observing rewards, and learns for how long to commit to the plan by following it without re-planning. Similar to GANs, STRAW could be used after the simulation learning stage so the agent copes with any discrepancies between the simulation and the real world.
 
Adaptive learning is a core characteristic to achieving strong AI \cite{lake2017building}. Several adaptive learning methods have been proposed which utilize prior knowledge \cite{levine2016end,chen2016learning,li2017deep}.  \cite{levine2016end} proposed to represent a particular optimization algorithm as a policy, and convergence rate as reward. \cite{chen2016learning,li2017deep} proposed to learn a flexible recurrent neural network (RNN) model to handle a family of RL tasks, to improve sample efficiency, learn new tasks in a few samples, and benefit from prior knowledge.

The notion of self-play is one of the biggest advancements of modern AI. AlphaGo AI is Deepmind's newest Go playing AI \cite{Silver2017Arxiv}, that learns, \textit{tabula rasa}, superhuman proficiency in challenging domains. Starting with the basic rules, they used self-play for the AI to learn strategies by playing against itself and storing efficient / rewarding moves. Fictitious Self-Play is a machine learning framework that implements fictitious play in a sample-based fashion \cite{heinrich2015fictitious}. The three strategies that are compared are: Learning by self-play, learning from playing against a fixed opponent, and learning from playing against a fixed opponent while learning from the opponent’s moves as well \cite{van2013reinforcement}. 

\section{Self-play Scenarios and Architectures}
Generalizing from training into real scenarios is not easy for self learning agents. This problem is known as the \emph{reality gap}~\cite{Jakobi1995ECAL}. In the initial stages of AI research, the training of self-learning agents included rules or limited scenarios where it can learn and improve upon competition against other introduced players. Interestingly, video games have emerged as one of  the main source of \emph{benchmark} environments for the training and testing of such agents, mostly due to its realistic, yet controlled approach to the real world, and the easy access to large amounts of data. For example, an AI agent is trained by playing with a perfect copy of itself without any supervision~\cite{Dosovitskiy2016Arxiv}. In this scenario, a set of basic rules of the game have been introduced at the beginning and the agent improves much faster using a vector of rewards instead of the \emph{classical} scalar quantity~\cite{sutton2011reinforcement}. 

Initially, self-play agents were trained to play board games (such as chess and go, among others)~\cite{Silver2017Arxiv} but it has now been successfully extended from the classic and \emph{simpler} Atari 2600 video games~\cite{Mnih2015Nature} to more complex first-person shooters: Doom~\cite{Kempka2016CIG}, Battlefield~\cite{Harmer2018Arxiv}, Quake~\cite{Beattie2016Arxiv}; Role Playing games\footnote{Including the Massively Multiplayer Online type of games.}: Warcraft~\cite{Andersen2017towards}; Real-Time Strategic Games: Starcraft~\cite{Vinyals2017Arxiv,Synnaeve2016Arxiv,Tian2017ANIPS} and more recently Multiplayer Online Battle Arena (Dota 2,~\cite{OpenaiDOTA2}). For a more comprehensive review see the work by \cite{Justesen2017Arxive}.

Motivation for studying self-play scenarios has increased in recent years mainly due to the advances in neural network architectures suitable to the reinforcement learning paradigm: 
DQN~\cite{Mnih2015Nature}, AC3~\cite{Mnih2016ICML}, 
DSR~\cite{Kulkarni2016Arxiv}, Dueling networks\cite{Wang2015Arxiv} among others as well as the development of powerful and accessible graphics processing unit (GPU) computing resources. This area of research is broadly known as Deep Reinforcement Learning~\cite{li2017deep,Mousavi2016deep,arulkumaran2017brief}.

The challenge of training self-playing agents in order to develop more complex policies inside realistic and highly specific or general environments remain as an open problem. Most of the recent developments tend to focus on very particular properties of the learning agent or the way that they interact with their surroundings. To address this issue, we identify two general mechanism that can be improved in order to design a better self-learning agent: self-play scenarios and self-learning architectures.

\begin{figure}
\begin{center}
\includegraphics[width=0.3\paperwidth]{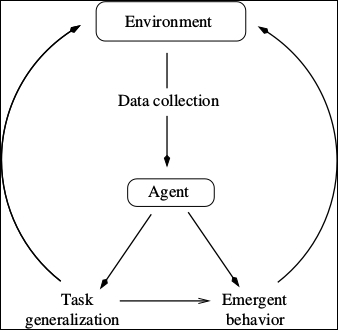}
\end{center}
\caption{Proposed general architecture for a self-learning agent interacting with its environment.}
\label{fig1}
\end{figure}

\subsection{Improving self-play scenarios and self-learning agents: closing the reality gap}

Constructing realistic self-play scenarios plays a fundamental role in training self-learning agents. Once an agent is immersed in a specific environment, we expect (independently of the self-learning architecture) that it will learn a set of policies accordingly to the received experiences\footnote{A learning agent can learn a set of policies, or will optimize the parameters of a given set of policy. New policies can emerge even without previous knowledge. The sum of the whole policies is called general policy.}. A problem which is widely understood, is that when agents learn from strict simulated scenarios, they may not be prepared for unexpected situations when the environment changes, such as a pigeon flying towards the sensor of a driverless car. Here, we propose a general scheme that uses the versatility of the video games or simulators as a source of synthetic data  and the wide array of capabilities of modern extended reality technologies, to enrich the properties of the real environment during the training of self-learning agents. An agent may learn independently, but the environment can be controlled to persuade the agent to learn a set of additional policies for unexpected scenarios. In addition to the enriched data, the self-learning agent may be trained using purely synthetic data. But the limitations of this method rely in the accuracy of the representation of the real scenarios. 

For the self-learning mechanism, we have identified three key steps which could improve the design of architectures for self-learning agents, which may improve policies both in terms of effectiveness and robustness. The three areas are: Data collection, task generalization and emergent behaviour (see Fig.~\ref{fig1}). For a given agent, in the first stage the agent will need to interact with the environment, possibly by accessing a data collection, then the agent should be able to generalize a set of given tasks and, simultaneously, new skills should emerge (independently or due to the task generalization). In the final stage, the emergent and the generalization skills interact with the environment to create a \emph{continuous} self-learning agent.  

\begin{figure}
\begin{center}
\includegraphics[width=0.3\paperwidth]{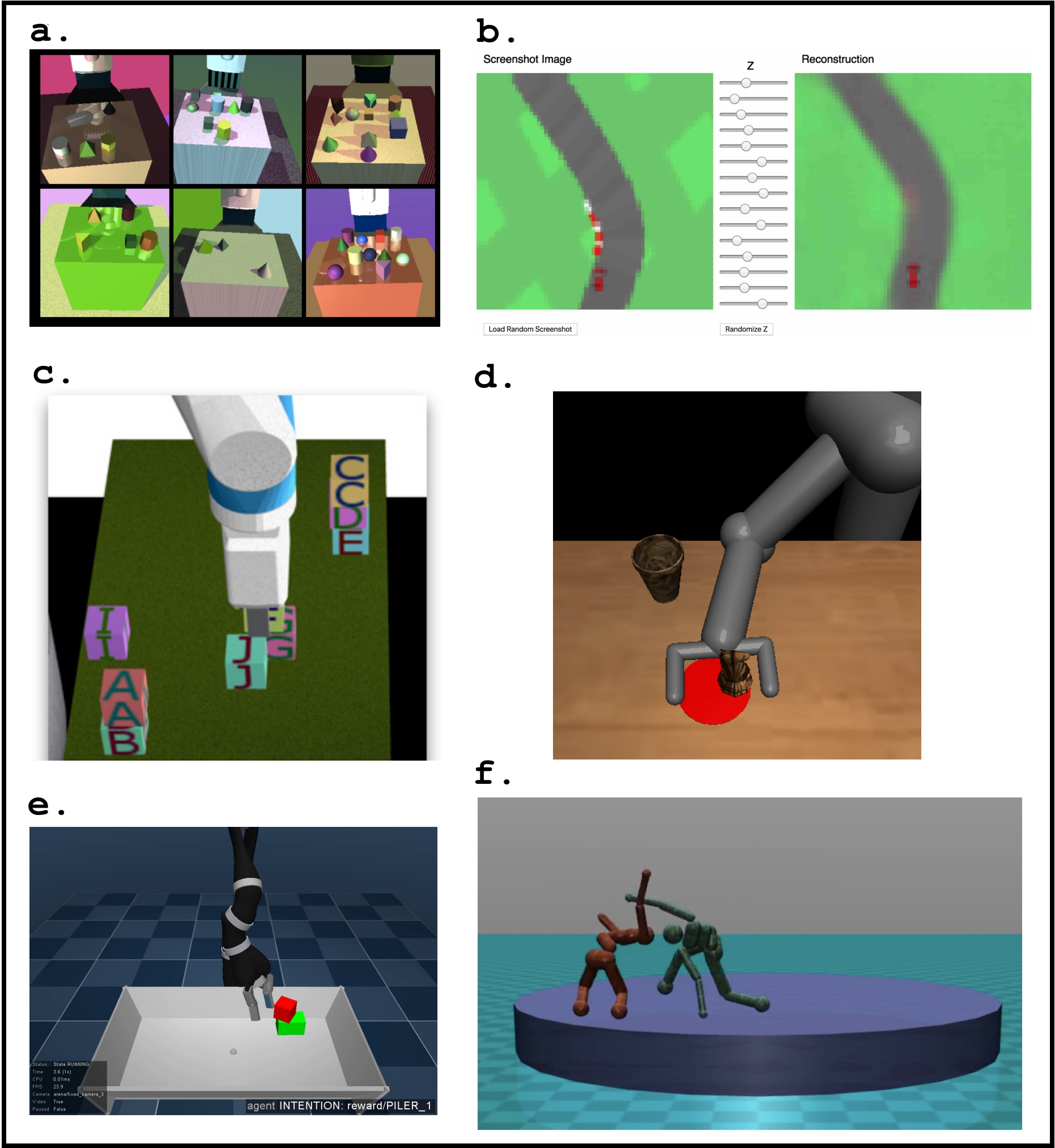}
\end{center}
\caption{Illustration of how different components from Figure~\ref{fig1} can be used on complete self-learning systems. \textbf{a}. Adapted from \cite{Tobin2017IROS}, \textbf{b}. Adapted from \cite{Ha2018Arxiv}, \textbf{c}. Adapted from \cite{Duan2017ANIPS}, \textbf{d}. Adapted from \cite{Finn2017Arxiv}, \textbf{e}. Adapted from \cite{Riedmiller2018Arxive} and \textbf{f}. Adapted from \cite{Bansal2017Arxiv}.
}
\label{fig2}
\end{figure} 

To illustrate the steps, we present a set of representative developments in the area of (deep) reinforcement learning. Note that this list is not comprehensive but the examples are presented to highlight their own specific properties then to introduce them in our general model (see Figure~\ref{fig2}). Each can be used as a building block inside a complete self-play scenario, for example, Figure~\ref{fig2}a shows a low-fidelity rendered images with random camera positions, lighting conditions, object positions, and non-realistic textures use to train a self-learning agent~\cite{Tobin2017IROS}; 
Figure~\ref{fig2}b shows an agent which uses a compressed representation of a real scenario to learn a set of policies which are successfully use back to the real environment~\cite{Ha2018Arxiv}. Figure~\ref{fig2}c shows an image of a robot used as a one-demonstration example during the training stage~\cite{Duan2017ANIPS}. Figure~\ref{fig2}d shows another image of a robot used during a training stage to teach a robot to place an object in a target position~\cite{Finn2017Arxiv},
Figure~\ref{fig2}e depicts an agent stacking two blocks, behaviour learn from sparse rewards~\cite{Riedmiller2018Arxive} and 
Figure~\ref{fig2}f illustrates one competitive environment where one the agents develops a new set skills~\cite{Bansal2017Arxiv}.


\begin{enumerate} 

\item Data collection:\\
\emph{Domain randomization (DR)}: 
DeepMind have recently shown how an agent can be trained on artificially generated scenarios. In their paper, the authors successfully transferred the knowledge from a neural network purely trained on low resolution rendered RGB images: 
\emph{domain randomization}~\cite{Tobin2017IROS}. This method can be extensively used for training agents in the case that the amount of data available is low or when the separation between the real and the train environment is immense.

\emph{World models (WM)}: 
Self-learning agents can be trained in a compressed spatial and temporal representation of real environments by \cite{Ha2018Arxiv}. This method is highly powerful because an agent can learn in a more compact or \emph{hallucinated} universe and then go back to the original environment exporting the set of learned abilities. One of the main advantages of this method is the possibility to perform a much faster and accurate \emph{in situ} training of the agents by using less demanding computational resources. 

\item Task generalization: 

\emph{One-shot imitation learning (OSIL)}:
the authors present an improved method that uses a meta-learning framework built upon the soft attention model~\cite{Bahdanau2014Arxive} named \emph{one-shot imitation learning}~\cite{Duan2017ANIPS}. Here, the agents are able to learn a new policy and solve a task after being exposed to a few demonstrations during the training stage.

\emph{One-shot visual imitation learning (OSVIL)}: Meta-imitation learning algorithm that teaches an agent to learn how to learn efficiently~\cite{Finn2017Arxiv}. In this work, a robot reuses past experiences and then upon a single demonstrations, it develops new skills. 


\item Emergent behaviour:

\emph{Scheduled auxiliary control (SAC)}: 
another research team from DeepMind introduced a new framework that allows agents to learn new and complex behaviours in presence of a stream of sparse rewards~\cite{Riedmiller2018Arxive}.

\emph{Multi-agent competition (MAC)}: 
a paper by one research team from OpenAI, the authors showed that multi-agents self-play on complex environments can produce behaviours which can be more complex than the environment itself~\cite{Bansal2017Arxiv}. The emergent skills can improve the capabilities of the agents upon unexpected  changes in the real environment.

\end{enumerate}

To summarize, Figure~\ref{fig1} illustrates how an agent can retrieves data from an environment and then generalizes to a specific task and simultaneously develops new abilities. The new skills can emerge independently or due to the task generalization process. In the final stage, the environment gets modified by the agent itself. A combination of such methods could be used to create more effective architectures for teaching self-learning agents. The novelty of this idea is that it builds upon the concept of modularity to create more complex architectures which develop new untrained behaviours. This notion  has been explored before to improve learning agents, for instance by synergistically combining two modelling methods such as type-based reasoning and policy reconstruction~\cite{albrecht2013game}. Interestingly, it has been stated recently that this is still a promising area under research~\cite{albrecht2018autonomous}. In addition, the proposed architecture establishes a general framework for the design of intelligent agents or rational agents searching to maximize their reward upon interaction with an external environment~\cite{russell2016artificial} via specific goals (task generalization and emergent behaviour). Once an architecture is defined, targeting the most adequate general policies are dependent on the external information gathered by the agent. In the data collection module, the main goal is to optimize the data used by the agent for learning. In the general case however, a proper design should include a non-static complete representation of the environment for exploring extensively all possible scenarios.

In the next section, we present a complete general design for a self-learning agent including an extra module for extending the description of the world experienced by the agent and then, upon an ad-hoc division and by using the concept of multi-agent~\cite{wooldridge2009introduction}. We discuss the application in the design of a central autonomous agent able provide via rational decision-making, strategies to tackle the possible consequences derived from changes on any stage of a specific industrial infrastructure.

\section{Proposed Design}

\begin{figure}
\begin{center}
\includegraphics[width=0.3\paperwidth]{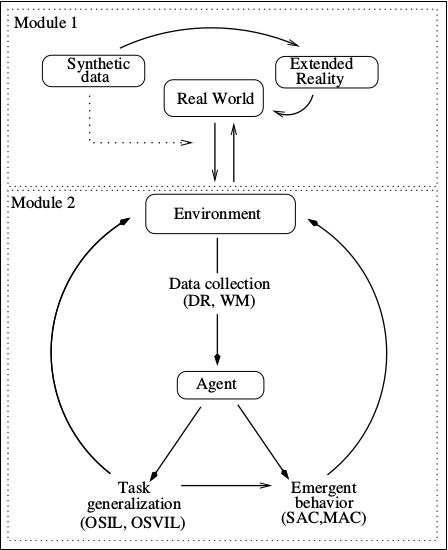}
\end{center}
\caption{Two module design of a general architecture for a self-learner agent interacting with its enriched or altered environment. }
\label{fig3}
\end{figure}


As already discussed, our goal of designing general self-play scenarios for teaching self-learning agents can be tackled by separating the data retrieved from the environment and the agent's self-learning architecture. In the spirit of the SMM (see Section \ref{research_methods}) we present a general scheme in which we divide the general architecture into two modules, in Module 1, the agent retrieves the data from its surroundings as a combination of information from the real world and synthetic data (or pure synthetic data), and in Module 2 (equivalent to the structure of the Fig~\ref{fig2}), the agent creates its final policies. The general scheme is depicted in the Fig~\ref{fig3}.

For a self-learning agent inside a specific self-learning scenario, there is no difference between synthetic or real data. Here we call real data the information extracted from physical world without any previous or further digital modifications. The agent uses exclusively the information, in terms of raw bytes, independently of the sensors that connect it with the environment. The use of synthetic data arises as a need to expose the self-learning agent to unexpected situations or conditions that allow it to create a set of optimal related policies. The representation of the real world can be done also via pure synthetic data, however the best case scenario is such where the synthetic data is used to extend the real world. The agent also can modify its own environment during the learning process, and if the whole process is fully done in a simulation environment, the use of engines that mimic the physical rules of the real world become necessary. The proposed general design can be encapsulated and used as a basic element in a more complex multi-agent based learning architecture~\cite{wooldridge2009introduction}. The communication among its moieties can be performed via the local information modules that represents their individual real worlds. Posteriorly, a central agent retrieves and updates its state providing a final decision. Several mathematical strategies can be applied to interchange the information, for instance, by defining scalar, vectorial or probabilistic representations.

In the next part we  aim to present an example or how we can build a minimal supra-modular architecture for the specific case of industrial management. We discuss the advantages and the characteristics for the modular parts of the design, as well as review (not exhaustively), the state-of-the-art developments for learning agents in decision making.

\subsection{Application in Autonomous Industrial Management}

The aforementioned improvements can be employed in specific applications using specific designs. In particular, by using the new and open simulation frameworks such as OpenAI Gym\footnote{https://github.com/openai/gym} or Dopamine\footnote{https://github.com/google/dopamine} among others. Here we discuss a particular application for industrial environments solving a managerial problem, however the developed ideas can be used for other purposes upon a clear definition of their parts. 

\begin{figure}
\begin{center}
\includegraphics[width=0.45\paperwidth]{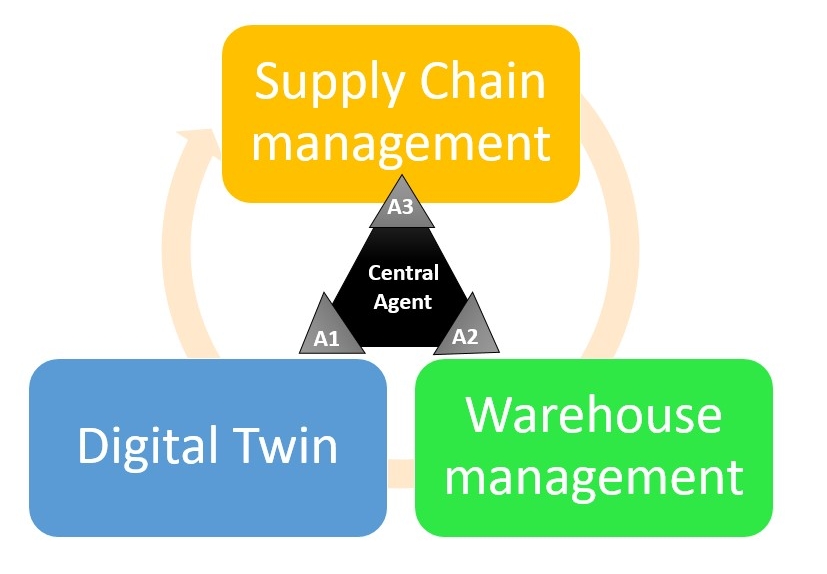}
\end{center}
\caption{Proposed architecture for a reinforcement learning agent in Autonomous Industrial Management (AIM) for adaptive industrial infrastructure.}
\label{industrial}
\end{figure}

In the industrial regime, decision-making is one of the key elements in the adaptive business intelligence discipline~\cite{michalewicz2006adaptive}. Despite the advances in the computational tools, still many relevant decisions are taken by real persons. Here, we propose a general model where self-learning agents can be trained to make decisions on industrial scale upon self-playing using extended reality training scenarios. We devised an strategy where an industrial infrastructure is divided ad-hoc in three independent sections managed by three independent self-learning agents (see Figure \ref{industrial}). Each agent replicates the physical process but it has the property to explore (and improve) in its own learning space based in the perturbation via self-learning mechanisms. Upon external perturbations in their environments, the final decision is done by a central agent which learns thanks to the information gathered for the independent agents. This Multi-Agent Reinforcement Learning (MARL) structure will bring a modularity structure that can be exploited in the industrial realm to improve and optimize the decision-making processes. Our example focuses mainly in the type of manufacturing industry, but the methodology can be extended easily to other industrial fields. Similar conceptual frameworks have been presented and described in the literature~\cite{Wooldridge2000, 10.1007/978-3-540-24620-6_15} and in some cases including full computational frameworks or software applied on specific industrial environments~\cite{castro2002towards,bordini2007programming} were described. The niche of industrial management is a fruitful source of academic research with the application of the cutting-edge methods for the design of intelligent systems. Our division in three different entities: Supply Chain Management (SCM), Warehouse or Inventory Management and  Digital Twin responds to the fact that a lot of work on modelling and creation of autonomous agents has been done in the last years independently in each area, therefore a low level description of an autonomous agent for industrial management is possible by following our proposed architecture (see Fig.~\ref{fig3}). In the next subsection we present a non exhaustive review of methods by highlighting some of the relevant strategies employed in the creation of autonomous systems.

\subsubsection{Supply Chain Management (SCM)}

The use of RL for optimization in the SCM has been discussed largely in the last decades~\cite{valluri2009reinforcement,zhang2018reference}, in particular, in the seminal works by \cite{barbuceanu1996coordinating} where the authors propose a multi-agent systems for the optimization of tasks in the SCM,  \cite{pontrandolfo2002global} that solves the multi-agent problem as a semi-Markov decision problem, \cite{stockheim2003reinforcement} that reduces the complexity by dividing the agents into three components (in a similar concept presented in this work), among others. Here, the self-learning agent must be able to decide about the different suppliers (external and internal) in coordination with the warehouse manager agent.

\subsubsection{Warehouse or Inventory Management}

Models for warehouse management have been developed in parallel to the state-of-the-art technological developments to achieve short and optimized responses in delivering goods \cite{van1999models} meanwhile an optimal stock level is maintained. In particular, it is worth to highlight two types of warehousing systems: automatic and automated warehousing systems. Several methods for self-learning agents in warehouses have shown to be effective in solving real problems, for instance, Stochastic Learning \cite{estanjini2011optimizing}, temporal difference Actor-Critic algorithm \cite{estanjini2012least} or more recently Deep Reinforcement Learning \cite{gijsbrechts2018can}. In general, the information can be controlled and retrieved via an optimized wireless network of sensors inside the warehouse \cite{golab2014data}. In this case, the self-learning agent must be able not only to gather the internal information about the location of the goods inside the warehouse but also  optimize the architecture of the network via self-play mechanisms.

\subsubsection{Digital Twin}

The use of digital tools in the industry has outgrown the level in which only individual applications were able to be modelled accurately for very specific topics, in the sixties, towards a complete simulation of the systems during the whole entire life cycle in modern times \cite{boschert2016digital}. The concept that includes a detailed model and description of a product, system or component in all its possible phases (product design, production system engineering, production planning, production execution, production intelligence and closed-loop optimization) \cite{rosen2015importance} is known as \emph{digital twin}\footnote{A term coined by NASA during the Apollo missions, but brought recently to the general public \cite{shafto2012modeling}.} In general terms, the digital twin is an extension of the model-based systems engineering (MBSE) concept \cite{wymore1993model}. The three important aspects of the digital twin have been discussed previously by other authors \cite{boschert2016digital,rosen2015importance,kritzinger2018digital}: modularity, connectivity and autonomy. These aspects can be linked with our proposed design in the Fig~\ref{fig3}. Our design can be extended into the MARL context.

\section{Conclusion \& Future Work}

The design of self-learning and self-play scenarios is still an area of fruitful development and research. Many critics have pointed out that AI research is limited by its own ideas~\cite{pearl2018book} as well as it's usefulness for industrial purposes. The creation and discussion of general architectures can open the door to new proposals, in particular, for the final emergence of the expected autonomous systems (see Section \ref{section21}). Despite a boom in the field during the last years, there are many open questions about how to enable self-learning agents to achieve specific tasks without any supervision. We propose to tackle this question by using our general architecture, which can be, in principle, limited by the modular developments and the availability of specific data sets or sensors.

In this work we grounded our position in autonomous systems by providing a definition of requirements and presented a general review of the relevant literature. We provided a proposal for a design of a general architecture for self-learning agents. Our design included two separate modules, one for the creation of the data and the second for the independent self-learning of an agent. We conclude by stating that the second module is, in general, divided into three stages, where each stage is in charge of accomplishing an independent task: data collection, task generalization and emergent behaviour. In very particular designs, generalization can influence emergent behaviours, but only in one direction. We expanded our findings by presenting a concrete example of our ideas in the industrial manufacturing sector. Here we suggest a division in three modular elements that feed a central agent for decision-making. Each module is able to explore and learn from all possible scenarios and available data, taking into account changes or disruptions in its own infrastructure and the information provided by the other modules. In the near future, we plan to present the first implementation of our ideas and publish the conclusions about the feasibility of our design.

\bibliographystyle{unsrt} 
\bibliography{main}

\end{document}